\documentclass[conference,letterpaper,final,twocolumn]{IEEEtran}
\usepackage[utf8]{inputenc}
\usepackage{amsmath}
\usepackage{amsthm}
\usepackage{url}

\usepackage{stfloats}
\usepackage{pstricks}
\usepackage{graphicx}

\newtheorem{theorem}{Theorem}
\newtheorem{proposition}{Proposition}

\newtheorem{definition}{Definition}

\title{Source Polarization}
\author{\IEEEauthorblockN{Erdal Ar{\i}kan}
\IEEEauthorblockA{
	Bilkent University\\ Ankara, Turkey}}
\author{\begin{tabular}{c}
Erdal Ar{\i}kan\\
Bilkent University, Ankara, Turkey\\
\end{tabular}}

\def\cX{\mathcal{X}}
\def\cY{\mathcal{Y}}

\begin{document}
\maketitle

\begin{abstract}
The notion of source polarization is introduced and investigated. This complements the earlier work on channel polarization. An application to Slepian-Wolf coding is also considered. The paper is restricted to the case of binary alphabets. Extension of results to non-binary alphabets is discussed briefly. 
\end{abstract}

\section{Introduction}

We introduce the notion of ``source polarization'' which complements ``channel polarization'' that was studied in \cite{ArikanIT2009}.
One immediate application of source polarization is the design of polar codes for lossless source coding. Lossless source coding using polar codes has already been considered extensively in the pioneering works \cite{HKU09} and \cite{KoradaThesis2009}, which reduced this problem to one of channel polarization using the duality between the two problems. The approach in this paper is direct and offers an alternative (primal) viewpoint. 

This paper is restricted mostly to binary memoryless sources. We indicate in the end briefly the possible generalizations to non-binary sources.

We use the notation of \cite{ArikanIT2009}. In particular, we write $u^N$ to denote a vector $(u_1,\ldots,u_N)$ and $u_i^j$ to denote the sub-vector $(u_i,\ldots,u_j)$ for any $1\le i\le j\le N$. If $j<i$, $u_i^j$ is the null vector. The logarithm is to the base 2 unless otherwise indicated. We write $X\sim \text{Ber}(p)$ to denote a Bernoulli random variable (RV) with values in $\{0,1\}$ and $P_X(1)=p$. The entropy $H(X)$ of such a RV is denoted sometimes as ${\cal H}(p)=-p\log p -(1-p)\log (1-p)$.

\section{Polarization of binary memoryless sources with side information}\label{section:Introduction}
Let $(X,Y)\sim P_{X,Y}$ be an arbitrary pair of random variables over $\cX\times \cY$ with $\cX =\{0,1\}$ and $\cY$ an arbitrary countable set. 
Throughout this section, we regard $(X,Y)$ as a memoryless source, with $X$ as the part to be compressed and $Y$ in the role of ``side-information'' about $X$.
We consider a sequence $\{(X_i,Y_i)\}_{i=1}^\infty$ of independent drawings from $(X,Y)$ and write $(X^N,Y^N)$ to denote the first $N$ elements of this sequence, for any integer $N\ge 1$.

\begin{figure}[thb]
\begin{center}
\psset{unit=0.5cm}
\psset{xunit=1,yunit=1}
\begin{pspicture}(0,0)(4,4)
\multirput[0,0]{0}(0,0)(0,2){2}{\psline[linecolor=black,linewidth=1pt]{-}(0,1)(3,1)}
\multirput[0,0]{0}(1.5,1)(0,4){1}{\psline[linecolor=black,linewidth=1pt]{-}(0,0)(0,2)}
\multirput[0,0]{0}(1.5,1)(0,4){1}{\pscircle[fillstyle=solid,fillcolor=white](0,2){0.4}\rput[c](0,2){$+$}\pscircle*(0,0){0.1}}

\rput[r](-0.5,1){$X_2$}
\rput[r](-0.5,3){$X_1$}
\rput[l](3.5,1){$U_2$}
\rput[l](3.5,3){$U_1=X_1 \oplus X_2$}

\end{pspicture}
\caption{Basic source transformation.}
\label{fig1}
\end{center}
\end{figure}
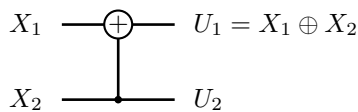

The basic idea of source polarization is contained in the transformation shown in Fig.~\ref{fig1}, where ``$\oplus$'' denotes addition mod-2. 
The operation $(X_1,X_2)\to (U_1,U_2)$ performed by the circuit preserves entropy, {\sl i.e.,}
\begin{align}
H(U_1,U_2|  Y_1,Y_2) & = H(X_1,X_2| Y_1,Y_2) \nonumber\\
& = 2H(X| Y)\label{EC},
\end{align}
but is polarizing in the sense that
\begin{align}
H(U_1| Y_1,Y_2) &\ge H(X| Y) \ge  H(U_2| Y_1,Y_2,U_1) \label{EP}.
\end{align}
It is easy to show that equalities hold here if and only if $H(X|Y)$ equals 0 or 1.
Thus, unless the entropies at the input of the circuit are already perfectly polarized, the entropies at the output will polarize further.

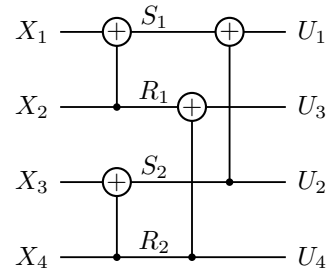
\begin{figure}[thb]
\begin{center}
\psset{unit=0.5cm}
\psset{xunit=1,yunit=1}
\begin{pspicture}(0,0)(7,8)
\multirput[0,0]{0}(0,0)(0,2){4}{\psline[linecolor=black,linewidth=0.7pt]{-}(0,1)(6,1)}
\multirput[0,0]{0}(1.5,1)(0,4){2}{\psline[linecolor=black,linewidth=0.7pt]{-}(0,0)(0,2)}
\multirput[0,0]{0}(1.5,1)(0,4){2}{\pscircle[fillstyle=solid,fillcolor=white](0,2){0.4}\rput[c](0,2){$+$}\pscircle*(0,0){0.1}}
\multirput[0,0]{0}(3.5,1)(0,8){1}{\psline[linecolor=black,linewidth=0.7pt]{-}(0,0)(0,4)}
\multirput[0,0]{0}(3.5,1)(0,8){1}{\pscircle[fillstyle=solid,fillcolor=white](0,4){0.4}\rput[c](0,4){$+$}\pscircle*(0,0){0.1}}
\multirput[0,0]{0}(4.5,3)(0,8){1}{\psline[linecolor=black,linewidth=0.7pt]{-}(0,0)(0,4)}
\multirput[0,0]{0}(4.5,3)(0,8){1}{\pscircle[fillstyle=solid,fillcolor=white](0,4){0.4}\rput[c](0,4){$+$}\pscircle*(0,0){0.1}}
\rput[r](-0.3,1){$X_4$}
\rput[r](-0.3,3){$X_3$}
\rput[r](-0.3,5){$X_2$}
\rput[r](-0.3,7){$X_1$}

\rput[l](6.3,1){$U_4$}
\rput[l](6.3,3){$U_2$}
\rput[l](6.3,5){$U_3$}
\rput[l](6.3,7){$U_1$}

\rput[c](2.5,1.4){$R_2$}
\rput[c](2.5,3.4){$S_2$}
\rput[c](2.5,5.4){$R_1$}
\rput[c](2.5,7.4){$S_1$}
\end{pspicture}
\normalsize
\caption{Four-by-four source transformation.}
\label{fig2}
\end{center}
\end{figure}

Figure~\ref{fig2} shows the recursive continuation of the construction to the case where four independent copies of $(X,Y)$ are processed. The entropy conservation law states that
$$
H(U^4| Y^4) = H(X^4| Y^4) = 4 H(X| Y).
$$
Using the chain rule, we may split the output entropy as
$$
H(U^4| Y^4) = \sum_{i=1}^4 H(U_i| Y^4,U^{i-1}).
$$
Note that the variables $U^4$ are assigned to the output terminals of the circuit in Fig.~\ref{fig2} in a shuffled order. This is motivated by the observation that, with this ordering, the pair $(U_1,U_2)$ is obtained from two i.i.d. RVs, namely, $(S_1,S_2)$, by the same two-by-two construction as in Fig.~\ref{fig1}. A similar remark applies to the relationship between $(U_3,U_4)$ and $(R_1,R_2)$.
These observations lead to the the following inequalities, which are special cases of those in \eqref{EP}.
\begin{align*}
H(U_1| Y^4) & \ge H(S_1| Y_1^2) \\
& = H(S_2| Y_3^4) \ge H(U_2| Y^4,U^1),
\end{align*}
\begin{align*}
H(U_3| Y^4,U^2) & \ge H(R_1| Y_1^2,S_1) \\
& = H(R_2| Y_3^4,S_2) \ge H(U_4| Y^4,U^3).
\end{align*}
There is no general inequality between $H(U_2| Y^4,U^1)$ and $H(U_3| Y^4,U^2)$.
The conclusion to be drawn is that polarization is enhanced further by repeating the basic construction.

For any $N=2^n$, $n\ge 1$, the general form of the source polarization transformation is defined algebraically as 
\begin{equation}\label{definition:GN}
G_N =\left[\begin{smallmatrix} 1 & 0 \\ 1 & 1 \end{smallmatrix}\right]^{\otimes n} B_N
\end{equation}
where ``$^{\otimes n}$'' denotes the $n$th Kronecker power and $B_N$ is the ``bit-reversal'' permutation (see \cite{ArikanIT2009}). 
It is easy to check that the transforms in Figures~\ref{fig1} and \ref{fig2} conform to $U^N = X^N G_N$.
The main result on source polarization for binary alphabets is the following. 
\begin{theorem}\label{theorem:binary}
Let $(X,Y)$ be a source as above. For any $N=2^n$, $n\ge 1$, let $U^N = X^N G_N.$
Then, for any $\delta \in (0,1)$, as $N\to \infty$,
\begin{equation*}
	\frac{\left|\bigl\{i\in[1,N]\colon
		H(U_i| Y^N,U^{i-1})\in(1-\delta,1]\bigr\}\right|
	}{N} \to H(X| Y)
\end{equation*}
and
\begin{equation*}
	\frac{\left|\bigl\{i\in[1,N]\colon
		H(U_i| Y^N,U^{i-1})\in[0,\delta)\bigr\}\right|
	}{N} \to 1-H(X| Y).
\end{equation*}
\end{theorem}

We omit the full proof but sketch the idea, which follows the proof of the channel polarization result in \cite{ArikanIT2009}. The first step is to define a tree random process for tracking the evolution of the conditional entropy terms $\{H(U_i| Y^N,U^{i-1})\}$. 
The analysis is aided by an accompanying supermartingale based on the source Bhattacharyya parameters. For the basic source $(X,Y)\sim P_{X,Y}$, this parameter is defined as 
$$
Z(X|Y) = 2\;\sum_{y} P_Y(y) \sqrt{P_{X|Y}(0|y)P_{X|Y}(1|y)}.
$$
The source Bhattacharyya parameters satisfy the following as they undergo the two-by-two polarization transformation. 
\begin{proposition}
Let $(X,Y)$ be a source as above, and $(X_1,Y_1)$ and $(X_2,Y_2)$ two independent drawings from $(X,Y)$.
Then,
$$
Z(X_1 \oplus X_2 | Y^2) \;\le\; 2Z(X| Y) - Z(X| Y)^2
$$
and
$$
Z(X_2 | Y^2,X_1 \oplus X_2) = Z(X| Y)^2.
$$
\end{proposition}
We omit the proof of this result since it is very similar to the proof of a similar inequality on channel Bhattacharyya parameters given in \cite{ArikanIT2009}.
Thus, we have the inequality
\begin{align*}
Z(U_1| Y^2) + Z(U_2| Y^2,U^1) & \le  2 Z(X| Y)
\end{align*}
which is the basis of the Bhattacharyya supermartingale. Convergence results about the Bhattacharyya supermartingale may be translated into similar results for the entropy martingale through the following pair of inequalities.
\begin{proposition}\label{proposition:ZvsH} For $(X,Y)$ a source as above, the following inequalities hold
\begin{align}
Z(X| Y)^2 & \le H(X| Y) \label{ineq:1}\\
H(X| Y) & \le \log(1+Z(X| Y))\label{ineq:2}.
\end{align}
Either both inequalities are strict or both hold with equality. For equality to hold, it is necessary and sufficient that $X$ conditioned on $Y$ is either deterministic or Ber$(\frac12)$.
\end{proposition}
The proof is given in the appendix.

These inequalities serve the purpose of showing that $H(X| Y)$ is near 0 or 1 if and only if $Z(X| Y)$ is near 0 or 1, respectively. Hence, the parameters $\{H(U_i| Y^N,U^{i-1})\}_{i=1}^N$ and $\{Z(U_i| Y^N,U^{i-1})\}_{i=1}^N$ polarize simultaneously. 

For coding theorems, it is important to have a rate of convergence result. 
\begin{definition}
Let $(X,Y)$ be a source as above, and let $R>0$. For $N=2^n$, $n\ge 1$, let $E_{X|Y}(N,R)$ denote a subset of $\{1,\ldots,N\}$ such that $|E_{X|Y}(N,R)|=\lceil NR\rceil$ and $Z(U_i|Y^N,U^{i-1}) \le Z(U_j|Y^N,U^{j-1})$ for all $i\in E_{X|Y}(N,R)$ and $j\notin E_{X|Y}(N,R)$. We refer to $E_{X|Y}(N,R)$ as a ``high-entropy'' (index) set of rate $R$ and block-length $N$. For the special case where $Y$ is absent or unavailable, we write $E_X(N,R)$ to denote the high-entropy set of $X$ only. When $N$ and $R$ are clear from the context, we simplify the notation by writing $E_{X|Y}$ or $E_X$.
\end{definition}

\begin{theorem}\label{thm:Rate}
Let $(X,Y)$ be a source as above and $R>H(X|Y)$ be fixed. Consider a sequence of high-entropy sets $\{E_{X|Y}(N,R):N=2^n, n\ge 1\}$. For any such sequence, any fixed $\beta < \frac 12$, and asymptotically in $N$, we have 
\begin{equation}
\label{eq:polar3}
\sum_{i\in E^c_{X|Y}(N,R)} Z(U_i|Y^N, U^{i-1}) =O(2^{-N^\beta}).
\end{equation}
\end{theorem}

We omit the proof, which is covered by the results of \cite{ArikanTelatarISIT2009}.

\section{Lossless source coding}\label{sec:Lossless}
Let $(X,Y)$ be a source as in the previous section and $(X^N,Y^N)$ denote an output block of length $N\ge 1$ produced by this source.
Shannon's lossless source coding theorem states that an encoder can compress $(X^N,Y^N)$ into a codeword of length roughly $NH(X| Y)$ bits so that a decoder observing the codeword and $Y^N$ can recover $X^N$ reliably, provided $N$ is sufficiently large. 
We now describe a method based on polarization that achieves this compression bound.
In the absence of any side information $Y^N$, the method given here is algorithmically identical to the source coding method proposed in \cite{HKU09} and \cite{KoradaThesis2009}; however, our viewpoint is different. Instead of reducing the source coding problem to a channel coding problem by exploiting a duality relationship between the two problems, we use direct arguments based solely on source polarization.

Fix $N=2^n$ for some $n\ge 1$. Fix $R>H(X|Y)$ and a high-entropy set $E_{X|Y}=E_{X|Y}(N,R)$.

{\sl Encoding:} Given a realization $X^N=x^N$, compute $u^N = x^N G_N$ and output $u_{E_{X|Y}}$ as the compressed word. 
(Note that the encoder does not require knowledge of the realization of $Y^N$ to implement this scheme.) 

{\sl Decoding:} Having received $u_{E_{X|Y}}$ and observed the realization $Y^N=y^N$, the decoder sequentially builds an estimate $\hat{u}^N$ of $u^N$ by the rule
\begin{align*}
\hat{u}_i & =\begin{cases} u_i & \text{if $i\in E_{X|Y}$}\\
0 & \text{if $i\in E_{X|Y}^c$ and $L_N^{(i)}(y^N,\hat{u}^{i-1})\ge 1$}\\
1 & \text{else}
\end{cases}
\end{align*}
where
$$
L_N^{(i)}(y^N,\hat{u}^{i-1}) = \frac{\Pr(U_i=0| Y^N=y^N,U^{i-1}=\hat{u}^{i-1})}{\Pr(U_i=1| Y^N=y^N,U^{i-1}=\hat{u}^{i-1})}
$$
is a likelihood ratio, which can be computed recursively using the formulas:
\begin{align*}
L_N&^{(2i-1)}(y^N,u^{2i-2}) \\
& =\frac{L_{N/2}^{(i)}(y^{N/2},u_{o}^{2i-2} \oplus u_{e}^{2i-2})L_{N/2}^{(i)}(y_{N/2+1}^N,u_{e}^{2i-2})+1}
{L_{N/2}^{(i)}(y^{N/2},u_{o}^{2i-2} \oplus u_{e}^{2i-2})+L_{N/2}^{(i)}(y_{N/2+1}^N,u_{e}^{2i-2})}
\end{align*}
and
\begin{align*}
L_N^{(2i)}&(y^N,u^{2i-1}) \\
& = L_{N/2}^{(i)}(y^{N/2},u_{o}^{2i-2}\oplus u_{e}^{2i-2})^{\delta_i}L_{N/2}^{(i)}(y_{N/2+1}^N,u_{e}^{2i-2})
\end{align*}
where $u_{o}^{2i-2}$ and $u_{e}^{2i-2}$ denote, respectively, the parts of $u^{2i-2}$ with odd and even indices, and $\delta_i$ equals 1 or -1 according to  $u_{2i-1}$ being 0 or 1, respectively.
Having constructed $\hat{u}^N$, the decoder outputs $\hat{x}^N = \hat{u}^N G_N^{\,-1}$ as the estimate of $x^N$. (It is easy to verify that $G_N^{\,-1} = G_N$.)

{\sl Performance:} 
The performance of the decoder is measured by the probability of error
\begin{align*}
P_e &= \Pr(\hat{U}^N \neq U^N) = \Pr(\hat{U}_{E^c_{X|Y}} \neq U_{E^c_{X|Y}}), 
\end{align*}
which can be upper-bounded by standard (union-bound) techniques as 
\begin{align}\label{eqn:PeBound}
P_e &\le \sum_{i\in E^c_{X|Y}(N,R)} Z(U_i| Y^N,U^{i-1}).
\end{align}

The following is a simple corollary to Theorem~\ref{thm:Rate} and \eqref{eqn:PeBound}.
\begin{theorem}
For any fixed $R>H(X|Y)$ and $\beta <\frac 12$, the probability of error for the above polar source coding method is bounded as $P_e = O(2^{-N^\beta})$. 
\end{theorem}

{\sl Complexity:}
The complexity of encoding and that of decoding are both $O(N\log N)$.

\section{Application to channel coding: Duality}
\begin{figure}[thb]
\begin{center}
\psset{unit=0.5cm}
\psset{xunit=1,yunit=1}
\begin{pspicture}(0,0)(10,2)
\psline[linecolor=black,linewidth=1pt]{-}(1,1)(2,1)
\psframe[linecolor=black,linewidth=1pt](2,0.2)(4,1.8)
\rput[c](3,1){$W^N$}
\psline[linecolor=black,linewidth=1pt]{-}(4,1)(6,1)
\psframe[linecolor=black,linewidth=1pt](6,0.2)(8,1.8)
\rput[c](7,1){$G_N$}
\psline[linecolor=black,linewidth=1pt]{-}(8,1)(9,1)
\pscircle[fillstyle=solid,fillcolor=white](1,1){0.1}
\rput[r](0.8,1){$Y^N$}
\pscircle[fillstyle=solid,fillcolor=white](5,1){0.1}
\rput[bc](5,1.6){$X^N$}
\pscircle[fillstyle=solid,fillcolor=white](9,1){0.1}
\rput[l](9.2,1){$U^N$}
\end{pspicture}
\caption{Channel coding.}
\label{fig3}
\end{center}
\end{figure}
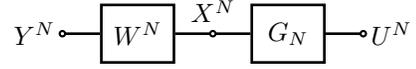

The above source coding scheme can be used to design a capacity-achieving code for any binary-input memoryless channel.
Let such a channel be defined by the transition probabilities $W(y|x)$, $x\in {\cal X}=\{0,1\}$ and $y\in {\cal Y}$. 
Consider the block coding scheme shown in Fig.~\ref{fig3}, where signals flow from right to left.
Here, $N=2^n$, $n\ge 1$, is the code block length; $U^N$ denotes the message vector, $X^N=U^N G_N$ the channel input vector, and $Y^N$ the channel output vector. Due to memorylessness, $W^N(y^N|x^N) = \prod_{i=1}^N W(y_i|x_i)$ for any $x^N\in {\cal X}^N$, $y^N\in {\cal Y}^N$.

We turn the triple $(U^N,X^N,Y^N)$ into a joint ensemble of random vectors by assigning the probabilities $\Pr(X^N=x^N)=2^{-N}$ for all $x^N\in \{0,1\}^N$. Under this assignment, $(X^N,Y^N)$ may be regarded as independent samples from a source $(X,Y)\sim Q(x)W(y|x)$ where $Q$ is the uniform distribution on $\{0,1\}$. We let $I(W)=I(X;Y)$ denote the symmetric channel capacity and fix $R< I(W)$. This implies that $1-R > H(X|Y)$. Let $E_{X|Y}=E_{X|Y}(N,1-R)$ denote a high-entropy set of rate $(1-R)$ for the source $(X,Y)$. The following coding scheme achieves reliable communication at rate $R$ over the channel $W$.

{\sl Encoding:} Prepare a binary source vector $U^N$ as follows. Pick the pattern $U_{E_{X|Y}}$ at random from the uniform distribution and make it available to the decoder ahead of the session. In each round, fill $U_{E^c_{X|Y}}$ with uniformly chosen data bits. (Thus, $\lfloor NR\rfloor$ bits are sent in each round, for a data transmission rate of roughly $R$.) Encode $U^N$ into a channel codeword by computing $X^N = U^N G_N$ and transmit $X^N$ over the channel $W$. 

{\sl Decoding:} Having received $Y^N$, use the source decoder of the previous section to produce an estimate $\hat{U}_{{E_{X|Y}^c}}$ of the data bits $U_{{E_{X|Y}^c}}$. 

{\sl Analysis:}
The error probability $\Pr(\hat{U}_{{E_{X|Y}^c}}\neq U_{{E_{X|Y}^c}})$ is bounded as $O(2^{-N^\beta})$ for any fixed $\beta < \frac 12$ since the source coding rate is $1-R>H(X|Y)$. The complexity of the scheme is bounded as $O(N\log N)$.  

{\sl Remark.\/} The above argument reduces the channel coding problem for achieving the symmetric capacity $I(W)$ of a binary-input channel $W$ to a source coding problem for a source $(X,Y)\sim QW$ where $Q$ is uniform on $\{0,1\}$. This reduction exploits the duality of the two problems. This dual approach provides an alternative proof of the channel coding results of \cite{ArikanIT2009}. It also complements the duality arguments in \cite{HKU09} and \cite{KoradaThesis2009},  where the source coding problem for a $\text{Ber}(p)$ source was reduced to a channel coding problem for a binary symmetric channel with cross-over probability $p$.

\section{Slepian-Wolf Coding}\label{section:SW}

The above source coding method can be easily extended to the Slepian-Wolf setting \cite{SlepianWolfIT1973}.
Suppose $\{(X_i,Y_i)\}_{i=1}^\infty$ are independent samples from a source $(X,Y)$ where both $X$ and $Y$ are binary RVs. In the Slepian-Wolf scenario, there are two encoders and one decoder. Fix a block-length $N=2^n$, $n\ge 1$, and rates $R_x$ and $R_y$ for the two encoders. Encoder 1 observes $X^N$ only and maps it to an integer $i_x\in [1,2^{NR_x}]$, encoder 2 observes $Y^N$ only and maps it to an integer $i_y \in [1,2^{NR_y}]$. The decoder in the system observes $(i_x,i_y)$ and tries to recover $(X^N,Y^N)$ with vanishing probability of error. The well-known Slepian-Wolf theorem states that this is possible provided $R_x\ge H(X| Y)$, $R_y \ge H(Y| X)$, and $R_x+R_y \ge H(X,Y)$. 

It is straightforward to design a polar coding scheme that achieves the corner point $(H(X| Y),H(Y))$ of the Slepian-Wolf rate region. 
Fix $R_y > H(Y)$ and $R_x > H(X|Y)$. For $N=2^n$, $n\ge 1$, consider a pair of high-entropy sets $E_Y=E_Y(N,R_y)$ and $E_{X|Y}=E_{X|Y}(N,R_x)$. 

{\sl Encoding:\/} Given a realization $X^N=x^N$, encoder 1 calculates $u^N=x^N G_N$ and sends $u_{E_{X|Y}}$ to the common decoder. 
Given a realization $Y^N=y^N$, encoder 2 calculates $v^N=y^N G_N$ and sends $v_{E_Y}$.

{\sl Decoding:\/} The decoder first applies the decoding algorithm of Section~\ref{sec:Lossless} to obtain an estimate $\hat{y}^N$ of $y^N$ from $v_{E_Y}$. Next, the decoder applies the same algorithm to obtain an estimate of $x^N$ using $\hat{y}^N$ (as a substitute for the actual realization $y^N$) and $u_{E_{X|Y}}$. 

We omit the analysis of this scheme since it essentially consists of two single-user source coding schemes of the type treated in Section~\ref{sec:Lossless}.

It is clear that polar coding can achieve all points of the Slepian-Wolf region by time-sharing between the corner points $(H(X),H(X|Y))$ and $(H(X|Y),H(Y))$.
 
We should remark that polar coding for Slepian-Wolf problem was first studied in \cite{HussamiKoradaUrbanke2009}, \cite{HKU09}, and \cite{KoradaThesis2009} under  the assumptions that $X,Y\sim \text{Ber}(\frac12)$, and $X\oplus Y\sim \text{Ber}(p)$.

The above approach to Slepian-Wolf coding reduces the problem to single-user source coding problems. A direct appoach would be to have each encoder apply polar transforms locally, with encoder 1 computing $U^N = X^N G_N$ and encoder 2 computing $V^N = Y^N G_N$. Preliminary analyses show that such local operations polarize $X_1^N$ and $Y_1^N$ not only individually but also in a joint sense. A detailed study of such schemes is left for future work.

\section{Polarization of non-binary memoryless sources}

\begin{theorem}\label{theorem:prime}
Let $X\sim P_X$ be a memoryless source over ${\cal X}=\{0,1,\ldots,q-1\}$ for some prime $q\ge 2$.
For $n\ge 1$ and $N=2^n$, let $X^N=(X_1,\ldots,X_N)$ be $N$ independent drawings from the source $X$. Let $U^N = X^N G_N$ where $G_N$ is as defined in \eqref{definition:GN} but the matrix operation is now carried out in GF($q$).
Then, the polarization limits in Theorem~\ref{theorem:binary} remain valid provided the entropy terms are calculated with respect to base-$q$ logarithms.
\end{theorem}

If $q$ is not prime, the theorem may fail. Consider $X$ over $\{0,1,2,3\}$ with $P_X(0)=P_X(2) =\frac12$. Then, it is straightforward to check that $U^N$ has the same distribution as $X^N$ for all $N$. 
On closer inspection, we realize that $X$ is actually a binary source under disguise. More precisely, $X$ is already polarized over $\{0,2\}$, which is a subfield of $GF(4)$, and vectors over this subfield are closed under multiplication by $G_N$.

The preceding example illustrates the difficulties in making a general statement regarding source polarization over arbitrary alphabets. If we introduce some randomness into the construction as in \cite{SasogluTelatarArikan2009}, it is possible to polarize sources over arbitrary alphabets, still maintaining the $O(N\log N)$ complexity of the construction.

\section*{Acknowledgment}
Helpful discussions with E. \c Sa\c so\u glu and S. B. Korada are gratefully acknowledged. This work was supported in part by The Scientific and
Technological Research Council of Turkey (T\"UB\.ITAK) under
contract no. 107E216, and in part by the European Commission
FP7 Network of Excellence NEWCOM++ (contract no. 216715).

\section{Appendix}

\subsection{Proof of Inequality~\eqref{ineq:1}}

First we prove that $Z(X)^2 \le H(X)$ for any $X\sim \text{Ber}(p)$ with equality if and only if $p\in \{0,\frac12,1\}$.
Let $F(p)= H(Z) - Z(X)^2 = -p\log_2(p) -(1-p)\log_2(1-p) -4p(1-p)$, and compute
$$
\frac{dF}{dp} = \frac{1}{\ln 2} \left[-\ln p + \ln (1-p) \right] - 4 + 8p,
$$
$$
\frac{d^2F}{dp^2} = \frac{1}{\ln 2} \left[-\frac{1}{p} - \frac{1}{1-p}\right] + 8,
$$
$$
\frac{d^3F}{dp^3} = \frac{1}{\ln 2} \left[\frac{1}{p^2} - \frac{1}{(1-p)^2}\right].
$$
Inspection of the third order derivative shows that $dF/dp$ is strictly convex for $p\in[0,\frac12)$ and strictly concave for $p\in(\frac12,1]$.
Thus, $dF/dp=0$ can have at most one solution in each interval $[0,\frac12)$ and $(\frac12,1]$.
Since $dF/dp=0$ at $p=\frac12$, the number of zeros of $dF/dp$ over $[0,1]$ is at most three.
Thus, $F(p)$ can have at most three zeros over $[0,1]$.
Since $F(p) =0$ for $p\in \{0,\frac12,1\}$, there can be no other zeros. 

Thus, for any pair of random variables $(X,Y)$ with $X$ binary, if we condition on $Y=y$, we have
$$
Z(X| Y=y)^2 \le H(X| Y=y).
$$
Averaging over $Y$, and by Jensen's inequality, we obtain \eqref{ineq:1}.

\subsection{Proof of Inequality~\eqref{ineq:2}}

Recall that the R\'enyi entropy of order $\alpha$ ($\alpha > 0$, $\alpha \neq 1$) for a RV $X$ is defined as
$$
H_\alpha(X) = \frac{1}{1-\alpha} \log \sum_x P_X(x)^\alpha
$$
and has the following properties \cite{Csiszar1995}.
\begin{itemize}
\item $H_\alpha(X)$ is strictly decreasing in $\alpha$ unless $P_X$ is uniform on its support $\text{Supp}(X) = \{x: P_X(x)>0\}$.
\item $H(X) = \lim_{\alpha \to 1} H_\alpha(X)$.
\end{itemize}
Now suppose $X\sim \text{Ber}(p)$ and note that
\begin{align*}
H_{\frac12}(X) & = \log \left[\sum_x \sqrt{P_X(x)}\right]^2 = \log (1 + Z(X)).
\end{align*}
Thus, we have
$$
H(X)  \le H_{\frac12}(X) = \log(1+Z(X)).
$$

It follows that, for any jointly distributed pair $(X,Y)$ with $X$ binary and any sample value $Y=y$
$$
H(X| Y=y)  \le \log(1+Z(X| Y=y)).
$$
Averaging over $Y$ and by Jensen's inequality, we obtain \eqref{ineq:2}.


\begin{thebibliography}{1}

\bibitem{ArikanIT2009}
E.~Ar{\i}kan, ``Channel polarization: A method for constructing
  capacity-achieving codes for symmetric binary-input memoryless channels,''
  {\em IEEE Trans. Inform. Theory}, vol.~55, pp.~3051--3073, July 2009.

\bibitem{HKU09}
N.~Hussami, S.~B. Korada, and R.~Urbanke, ``Performance of polar codes for
  channel and source coding,'' in {\em Proc. 2009 IEEE Int. Symp. Inform.
  Theory}, (Seoul, South Korea), pp.~1488--1492, 28 June - 3 July 2009.

\bibitem{KoradaThesis2009}
S.~B. Korada, {\em Polar codes for channel and source coding}.
\newblock PhD thesis, EPFL, Lausanne, 2009.

\bibitem{ArikanTelatarISIT2009}
E.~Ar{\i}kan and E.~Telatar, ``On the rate of channel polarization,'' in {\em
  Proc. 2009 IEEE Int. Symp. Inform. Theory}, (Seoul, South Korea),
  pp.~1493--1495, 28 June - 3 July 2009.

\bibitem{SlepianWolfIT1973}
D.~Slepian and J.~Wolf, ``Noiseless coding of correlated information sources,''
  {\em IEEE Trans. Inform. Theory}, vol.~19, pp.~471--480, Jul. 1973.

\bibitem{HussamiKoradaUrbanke2009}
N.~Hussami, S.~B. Korada, and R.~L. Urbanke, ``Polar codes for channel and
  source coding.'' \url{http://arxiv.org/abs/0901.2370}, 2009.

\bibitem{SasogluTelatarArikan2009}
E.~{\c S}a{\c s}o{\u g}lu, E.~Telatar, and E.~Ar{\i}kan, ``Polarization for
  arbitrary discrete memoryless channels,'' {\em CoRR}, vol.~abs/0908.0302,
  2009.

\bibitem{Csiszar1995}
I.~Csisz\'ar, ``Generalized cutoff rates and {R}\'enyi's information
  measures,'' {\em IEEE Trans. Inform. Theory}, vol.~41, pp.~26--34, Jan. 1995.

\end{thebibliography}
\end{document}